\documentclass[8.5pt,twoside,twocolumn]{article}
\usepackage[super,sort&compress,comma]{natbib} 
\usepackage{mhchem}
\usepackage{times,mathptmx}
\usepackage{sectsty}
\usepackage{balance} 

\usepackage{graphicx} 
\usepackage{lastpage}
\usepackage{lineno}
\usepackage[format=plain,justification=raggedright,singlelinecheck=false,font=small,labelfont=bf,labelsep=space]{caption} 
\usepackage{fancyhdr}
\usepackage{color}
\usepackage[usenames,dvipsnames,svgnames,table]{xcolor}

\newcommand{\kp}[1]{{ \color{black} #1}}
\newcommand{\red}[1]{{\color{black} #1}}
\begin{document}

\thispagestyle{plain}
\fancypagestyle{plain}{
\renewcommand{\headrulewidth}{1pt}}
\renewcommand{\thefootnote}{\fnsymbol{footnote}}
\renewcommand\footnoterule{\vspace*{1pt}%
\hrule width 3.4in height 0.4pt \vspace*{5pt}} 
\setcounter{secnumdepth}{5}

\makeatletter 
\def\subsubsection{\@startsection{subsubsection}{3}{10pt}{-1.25ex plus -1ex minus -.1ex}{0ex plus 0ex}{\normalsize\bf}} 
\def\paragraph{\@startsection{paragraph}{4}{10pt}{-1.25ex plus -1ex minus -.1ex}{0ex plus 0ex}{\normalsize\textit}} 
\renewcommand\@biblabel[1]{#1}            
\renewcommand\@makefntext[1]%
{\noindent\makebox[0pt][r]{\@thefnmark\,}#1}
\makeatother 
\renewcommand{\figurename}{\small{Fig.}~}
\sectionfont{\large}
\subsectionfont{\normalsize} 

\fancyfoot{}
\fancyfoot[RO]{\footnotesize{\sffamily{1--\pageref{LastPage} ~\textbar  \hspace{2pt}\thepage}}}
\fancyfoot[LE]{\footnotesize{\sffamily{\thepage~\textbar\hspace{3.45cm} 1--\pageref{LastPage}}}}
\fancyhead{}
\renewcommand{\headrulewidth}{1pt} 
\renewcommand{\footrulewidth}{1pt}
\setlength{\arrayrulewidth}{1pt}
\setlength{\columnsep}{6.5mm}
\setlength\bibsep{1pt}

\twocolumn[
  \begin{@twocolumnfalse}
\noindent\LARGE{\textbf{Dynamics, crystallization and structures in colloid spin coating}}
\vspace{0.6cm}

\noindent\large{\textbf{Moorthi Pichumani,\textit{$^{a\ddag}$} Payam Bagheri,\textit{$^{b}$} Kristin M. Poduska,\textit{$^b$} Wenceslao {Gonz\'{a}lez-Vi\~{n}as},\textit{$^a$} and Anand Yethiraj$^{\ast, b}$}} \vspace{0.5cm}

\noindent\textit{\small{\textbf{Received Xth XXXXXXXXXX 20XX, Accepted Xth XXXXXXXXX 20XX\newline
First published on the web Xth XXXXXXXXXX 200X}}}

\noindent \textbf{\small{DOI: 10.1039/b000000x}}
\vspace{0.6cm}

\noindent \normalsize 
Spin coating is an out-of-equilibrium technique for producing polymer films and colloidal crystals quickly and reproducibly. In this review, we present an overview of theoretical and experimental studies of the spin coating of colloidal suspensions. The dynamics of the spin coating process is discussed first, and we present insights from both theory and experiment. A key difference between spin coating with polymer solutions and with monodisperse colloidal suspensions is the emergence of long range (centimeter scale) orientational correlations in the latter. We discuss experiments in different physical regimes that shed light on the many unusual partially-ordered structures that have long-range orientational order, but no long-range translational order. The nature of these structures can be tailored by adding electric or magnetic fields during the spin coating procedure. These partially-ordered structures can be considered as model systems for studying the fundamentals of poorly crystalline and defect-rich solids, and they can also serve as templates for patterned and/or porous optical and magnetic materials.

\vspace{0.5cm}

 \end{@twocolumnfalse}
]

\section{Introduction}

One of the earliest forms of production technology is the potter's wheel: a machine, in use for more than 5000 years, that produces axially symmetric ceramic pottery in a manner that is rapid and reproducible. This ancient concept of axial symmetry was rediscovered several decades ago to make thin films of paint, varnish and asphalt \cite{Walker1922, Kleinschmidt1953}. It is now applied routinely as an inexpensive batch production technology to make uniform thin polymer photoresist films for microelectronics applications \cite{Norrman2005}.

Colloidal films have one characteristic that gives them very wide scope in materials applications: they can be used as templates whose 3-dimensional pattern is then transferred to make materials of desirable chemistry \cite{Mann1996} for a diverse range of optical \cite{Wijnhoven1998}, magnetic \cite{Bartlett2003}, and surface wetting applications. Single-layer films can be used as shadow templates to coat surfaces with a pattern that is the inverse of the colloidal film. Multilayer films can be used as templates in conjunction with electrochemical surface coating to make thicker inverse structures \cite{Wijnhoven1998,Bartlett2003,Arcos2008}. In addition, colloidal films can be used directly to produce superhydrophobic surfaces \cite{Joung2011} and arrays for biological sensing \cite{Yang2006}. 

The focus of this review is the physics involved in spin coating colloidal suspensions, which is an out-of-equilibrium technique to make substrate-supported colloidal crystals with thicknesses ranging from a single layer to tens of layers. 
There is an intrinsic conceptual problem in making uniform crystalline colloidal films $via$ spin coating. While the axial symmetry of the spinning does not affect the degree of disorder in spin coated polymer films, the axial symmetry has a profound effect on the crystallization of colloidal suspensions. Therefore, the development of colloidal spin coating technology requires the development of strategies to control structure formation.

Most techniques that are used to produce colloidal crystals aim to produce large-scale single crystals, rather than polycrystals with the axial symmetry enforced by the spin coating process. The dip coating method \cite{Dimitrov1996} is inspired by the Langmuir-Blodgett technique \cite{blodgett1935}, and involves the slow extraction of a substrate from a colloidal suspension. If the liquid phase of the suspension is evaporated instead, the technique is called vertical deposition\cite{Jiang1999} (or controlled drying or convective assembly). \red{ Convective assembly has been demonstrated to allow control over symmetries and film thickness\cite{Ishii2005,Prevo2007,Merlin2012}.} These kinds of techniques are slow and prone to pattern-forming instabilities,\cite{Ghosh2007} which are undesirable if the goal is to make single crystals. Roll coating (or ``doctor blade coating'' \cite{Yang2010, Yang2010a}) has also been used to make large macroporous crystalline films. However, precise control of film thickness has not yet been demonstrated. This is a deficiency because single layer films are essential for shadow template applications, while thicker multilayer films are necessary for photonic applications. 

\begin{figure*}[hbt]
\centering
\includegraphics[width=180mm]{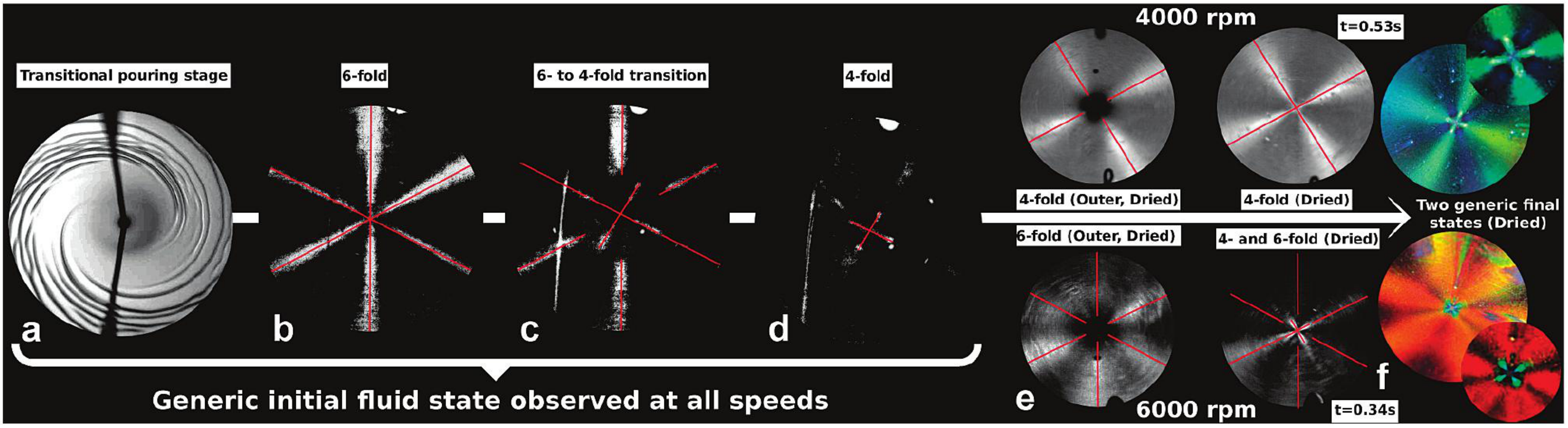} 
\caption{Symmetry transitions during spin coating\cite{Giuliani2010}. From a) to d), consecutive stages of the drying of a colloidal suspension during the spin coating process are shown. Disorder-order (emergence of 6-fold symmetry), order-order (6-fold to 4-fold symmetry), order-disorder (disappearance of 4-fold arms) transitions are seen in b, c and d respectively; these symmetry transitions are observed for all rotation rates. e) During the late stage of drying, either 4-fold or 6-fold symmetry emerges, depending on the experimental parameters. f) Two different configurations of the dried state, exhibiting 4-fold (top) and 6-fold (bottom) symmetry respectively. (Reprinted from J. Phys. Chem. Lett. 2010, 1, 1481–1486. Copyright 2010 American Chemical Society.)}
\label{fig:spin coating3}
\end{figure*}
A cursory survey of the colloid spin coating literature, reaching back as far as 1922, \cite{Walker1922} would leave a reader uncertain about what aspects of colloidal crystallization can be controlled with this out-of-equilibrium technique, and how viable it might ultimately be for producing crystals or templates that are useful for technological applications. Until now there is no unifying answer in the literature. \kp{In this context, this review highlights} that there are significant and interesting order-order and order-disorder transitions observed in the dynamics of spin coating \cite{Giuliani2010}. An understanding of the mechanisms of these phase transitions could lead to strategies for making more controllably ordered or disordered films. \kp{This review also} demonstrates that colloid spin coating offers a remarkably reproducible way to study crystallization in systems far from equilibrium, and that new techniques associated with spin coating \cite{bartlett_langmuir12,pichumanimag2011,Pichumani2012} hold significant promise for advancing the field further.

\section{Symmetry transitions during spin coating}
The most spectacular aspect of spin coating with monodisperse colloidal suspensions is the emergence
of symmetric structural colors within tens of milliseconds. While the following sections are chronologically faithful accounts of the dynamics, structure, and symmetries in spin coated colloidal suspensions, we present first a high-speed microscopy experiment of colloid spin coating, carried out by Giuliani \textit{et al.}\ \cite{Giuliani2010}, that will help to motivate this discussion.

In the experiments by Giuliani \textit{et al.}, silica microspheres (with micrometer-range diameters) were suspended in methyl ethyl ketone, a volatile solvent, and the suspension was dropped onto a spinning substrate. To view structural colors, as with viewing a rainbow, the most effective geometry is with lighting from behind,\textit{ i.e.} with the light source next to or behind the camera. The transient dynamics follow a very repeatable sequence. First, when the suspension is pipetted onto the substrate, the fluid spirals outward (Fig.~\ref{fig:spin coating3}a). When the spiral disappears, six symmetric reflection arms (with bright colors in the visible spectrum) appear (Fig.~\ref{fig:spin coating3}b). This order prevails for hundreds of milliseconds, following which there is a transition to 4-fold symmetry (Fig.~\ref{fig:spin coating3}c). This 4-arm pattern lasts between 10 and 100 ms, after which it shrinks towards the center leaving a dark film with no symmetric reflections (Fig.~\ref{fig:spin coating3}d). Up to this point in the spin coating process, all phenomena described above are qualitatively independent of the substrate rotation rate. Finally, the suspension dries, with the drying front starting from the edges and propagating inwards to give rise to either bright 4-fold or 6-fold reflections, depending on the rotation rate (Fig.~\ref{fig:spin coating3}e,f). Thus, there are two stages in the spin coating process: the dynamical stage (Fig.~\ref{fig:spin coating3}a-d), and the drying stage (Fig.~\ref{fig:spin coating3}e,f). 

\red{In what follows, we will see that different experimental conditions can re-create different subsets of the dynamical \kp{phenomena (Section 3) as well as the drying phenomena (Section 4), as illustrated in Fig.~\ref{fig:spin coating3}.}
For example, the dynamics in stages b and c of Fig.~\ref{fig:spin coating3} correspond closely with experiments carried out in non-volatile solvents\cite{Jiang1999,YuLing2007thesis,Shereda2008,Vermolen2008thesis}, while stages e and f correspond closely to experiments carried out with volatile solvents\cite{Mihi2006, Arcos2008}. In contrast, the early stage (a) of the dynamics has not yet been studied in detail.
}
\section{Dynamics}

Spin coating is remarkably simple to execute. It is, nevertheless, not trivial to identify the key fundamental parameters that govern the dynamics of colloidal crystal formation during the spin coating process. 

The dynamics of spin coating involves the evolution of a fluid phase in contact with a rotating disk. Consequently, it is a problem best considered by fluid mechanics methods. Although reported experiments in colloid spin coating began almost a century ago with Walker and Thompson \cite{Walker1922}, a complete theory has not yet emerged. However, the complexity of viscous fluids over a rotating disk-like propeller has been addressed in models since the time of von K\'arm\'an \cite{Karman1921,Cochran1934,Rehg1992}. Common approximations to simplify the problem of modeling colloidal spin coating are to neglect the effects of evaporation (including concentration changes), the presence of a gas phase, the possibility of non-Newtonian behavior, and the particulate character of the suspension.

As a first approximation, it is possible to use models of pure fluids or of molecular-scale homogeneous mixtures. The first model of Emslie, Bonner and Peck \cite{Emslie1958}  considered spinning a viscous, non-volatile fluid. They used realistic assumptions of axial symmetry, a thickness $z$ that is orders of magnitude smaller than the width of the spinning disk (enabling a lubrication approximation), and an incompressible flow. With this, the Navier-Stokes equations lead to
\begin{equation}
\frac{\partial u_r}{\partial t} = -\omega^2 r = \frac{\eta}{\rho} \frac{\partial^2 u_r}{\partial z^2} 
\label{timederiv}
\end{equation}
where $\eta$ and $\rho$ are the viscosity and density of the fluid, $u_r$ is the radial component of the velocity, and the acceleration has been rewritten in terms of the angular velocity $\omega$.

From Eq.\ \ref{timederiv}, the Emslie model leads to a quasi-linear first order partial differential equation for the thickness of the fluid layer $h$,

\begin{equation}
\frac{\partial h}{\partial t} + \frac{\omega^{2}r}{\nu}h^{2}\frac{\partial h}{\partial r} = -\frac{2\omega^{2}}{3\nu}h^{3} 
\label{emslie}
\end{equation}

\noindent where $\nu$ is the kinematic viscosity. Eq. \ref{emslie} can be easily integrated by the method of characteristics to give an implicit solution that depends on the distance from the center of rotation $r$,

\begin{equation}
h(r,t)=\frac{h_{0}\left[ r\left( 1-\sigma t h^{2} \right)^{3/4} \right] }{\sqrt{1+\sigma t\left\{ h_{0}\left[ r\left( 1-\sigma t h^{2} \right)^{3/4} \right] \right\}^{2}}}
\label{emsliesol}
\end{equation}

\noindent where $h_{0}(x)=h(x,0)$ represents the initial condition for the thickness and $\sigma= (4\omega^{2})/(3\nu)$. Emslie proved that, under some circumstances, simple fluids have thicknesses that become uniform (planarize) relatively fast during spin coating \cite{Emslie1958}.

\begin{figure}[ht]
\centering
\includegraphics[scale=1.3]{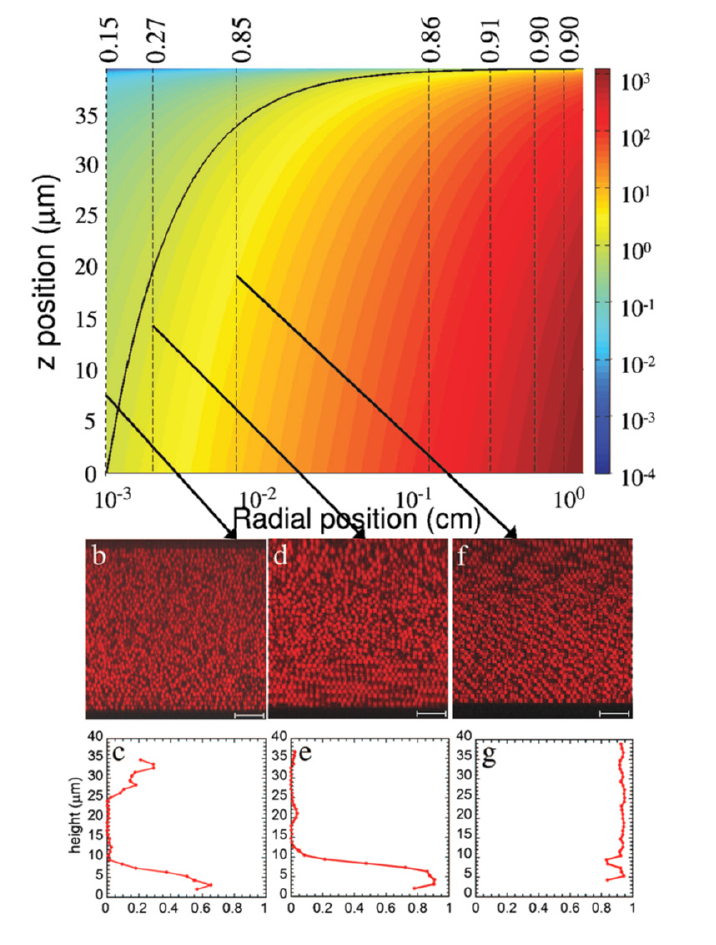}
\caption{$Z$ position (black curve, left) and Peclet number (color coded, right) plotted against radial position (bottom) and degree of crystallinity (top). When the Peclet number rises above a critical value (the yellow region), increasingly high degrees of crystallinity $\mathrm{(b,c) \rightarrow (d,e) \rightarrow (f,g)}$ appear. Reprinted
 with permission from ref. 23. Copyright (2008) by the American Physical Society. http://http://prl.aps.org/abstract/PRL/v101/i3/e038301 \cite{Shereda2008}}
\label{fig:spin coating1}
\end{figure}

One might reasonably expect that all bets are off when considering particulate suspensions through simple continuum theory, since it is unclear if this theory should still be relevant for films with thicknesses that are tens of particle diameters or less. Instead, there is a remarkably simply stated result for the character of crystallinity in a spinning sediment\cite{Shereda2008}. Eq.\ \ref{timederiv} can be rewritten in terms of a physically measurable quantity such as the shear stress $\tau_{rz} = - \eta \partial u_r/\partial z$ to give
\begin{equation}
\frac{\partial \tau_{rz}}{\partial z} = \rho \omega^2 r
\end{equation}
Integrating over a film of thickness $h$, one obtains a shear stress profile\cite{Shereda2008}
\begin{equation}
\sigma_{rz} = \rho \omega^2 r (h - z)
\end{equation}
The key result for non-volatile colloid spin coating, shown by Shereda \textit{et al.}, was that the local degree of crystallinity could be closely correlated with the local stress. Rewritten in terms of a Peclet number $Pe = \sigma_{rz} a^3/k_B T$, they showed that crystalline domains emerge when the Peclet number exceeds a critical value (with a magnitude of order unity).
\red{The results of Fig.~\ref{fig:spin coating1} are expected to carry over to the early stages of spin coating in volatile solvents, before the onset of the regime where evaporation dominates. The final stage of the evaporative spin-coating process, on the other hand, is likely to be dominated by restructuring of colloids at the solvent-air interface. }

For spin coating with a volatile solvent, Meyerhofer\cite{Meyerhofer1978} allowed for the thickness to change additionally due to solvent evaporation by including a correction, $E$, to the Emslie model:

\begin{equation}
\frac{dL}{dt} = -(1-c)\frac{2\omega^2 h^3}{3\nu}-E
\end{equation} 
where $dL/dt$ is the derivative of the volume of the solvent per unit area (assuming a homogeneous mixture), $c$ is the concentration of the solid (v/v), $h$ is the thickness and $\nu$ is the kinematic viscosity. This approximation assumes that the spin coating process consists of two different stages: flow dominates initially, and solvent evaporation dominates near the end. Later, Cregan {\it et al.} \cite{Cregan2007} generalized this result by considering solvent evaporation in both stages. Rotation-rate dependent diffusion and advection of the solvent, in both vapor and liquid phases, could affect evaporation rate\cite{Meyerhofer1978, Rehg1992, Cregan2007, Giuliani2010, Pichumani2012}, thereby yielding different thicknesses for the deposited layers.

\begin{figure}[h]
\centering
\includegraphics[scale=1]{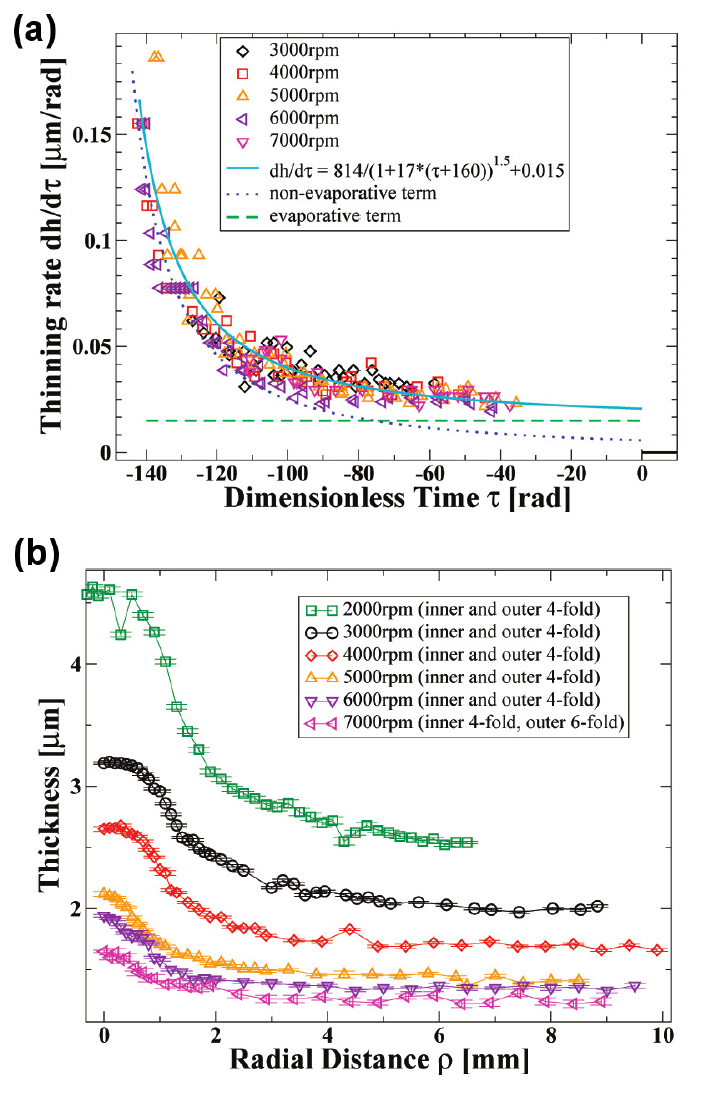} 
\caption{(a) Thickness as a function of time. Time is expressed in non-dimensional units as $\tau =  \omega (t - t_{\mathrm{dry}})$; zero time refers to the instant when the solvent completely dries. Thus all times of interest are negative dimensionless numbers. The thinning rate has a universal form that can be fit well to a simple model\cite{Giuliani2010}. (b) Thickness as a function of distance from the center of spinning shows a decrease in thickness as a function of radial distance, in contrast to assumption of planarization widely used in models of spin coating\cite{Giuliani2010}. (Reprinted from J. Phys. Chem. Lett. 2010, 1, 1481–1486. Copyright 2010 American Chemical Society.)}
\label{fig:spin coating2}
\end{figure}	
Due to drainage and evaporation of the solvent, film thickness (and its rate of thinning) changes continuously as a function of time. Experiments that yield the thinning rate can therefore be used to evaluate the validity of different models for spin coating. High-speed microscopy studies of colloid spin coating carried out by Giuliani \textit{et al.}\ \cite{Giuliani2010} also used a specular reflection geometry, where they observed thickness fringes instead of symmetric arms. By tracking interference fringes, the authors obtained a thinning rate $dh/d \tau$, where $\tau =  \omega (t - t_{\mathrm{dry}})$ is a dimensionless time, with zero time referring to the time that the solvent completely dries. This work found, remarkably, that the thinning rate followed a universal curve, Fig.\ \ref{fig:spin coating2}(a), for all spinning rates $\omega$.
\red{In spin-coating with non-volatile solvents, we expect that the thinning rate follows the dotted line in the diagram of Fig.~\ref{fig:spin coating3}(a). For volatile solvents, the additional physics in the thinning rate appears simply to be contained in a constant evaporative term.}

When there is a free surface with no lateral confinement, as in the case of spin coating, the fluid depth is higher at the center of rotation initially, but tends to planarize, i.e. become more uniform in height, over time. Some models predict a leveling time for simple fluids\cite{Bornside1991}.
An important difference between these experimental results and the simple models is the fact that the sediments do not become planar within the time span of the experiments\cite{Scriven1988,Giuliani2010}, but are instead significantly thicker near the center; the example in Fig.~\ref{fig:spin coating2}(b) displays atomic force microscopy height profiles obtained as a function of radial position for different rotation rates. %
Non-planarization has been linked to the non-Newtonian character of the colloids\cite{Scriven1988,Shereda2008}, since shear stresses will increase further from the center of rotation. Colloids suspended in volatile solvents present an even more complex situation, since the viscosity (and usually the non-Newtonian character) increases while spinning a suspension that becomes progressively more concentrated. To improve spin coating models, it will be essential to understand why planarization is not very important with respect to thinning rates.

Finally, it is worth mentioning that the relevance of flow instabilities (Ekman-like spirals\cite{Wahal1993}, hydraulic jump and fingering instabilities\cite{Melo1989}, the appearance of striations\cite{Du1995,Haas2002} and comet-driven instabilities, for example) have yet to to be explored in depth in the context of colloidal spin coating. 

\section{Structure and symmetries}

\subsection{Studies with non-volatile and volatile solvents}
\begin{figure}[ht]
\centering
\includegraphics[width=80mm]{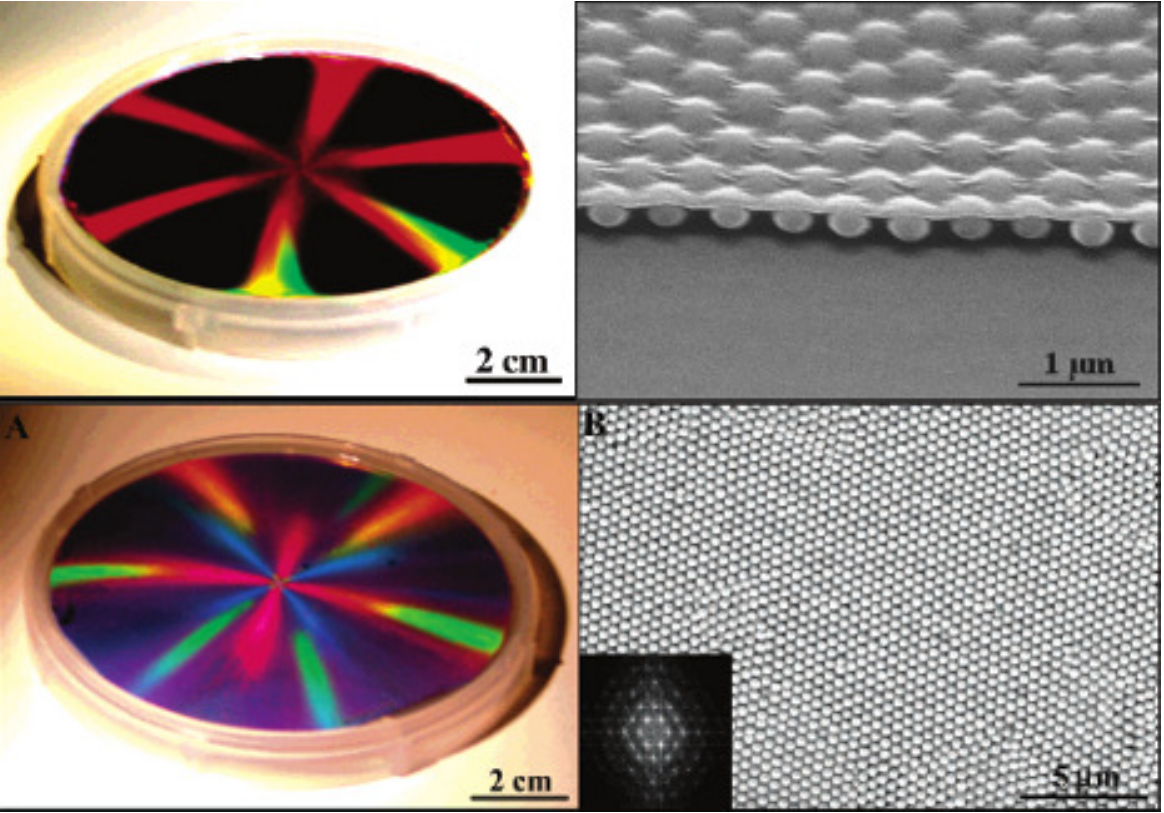} 
\caption{Spin coating in non-volatile solvents.
Top and bottom panels, left: cm-scale colloidal crystal exhibiting 6-fold symmetry and mixed 6- and 4-fold symmetry~\cite{Jiang2004}.(Reprinted from J. AM. CHEM. SOC. 2004, 126, 13778-13786. Copyright 2004 American Chemical Society.) Top, right: A single-layer colloidal film~\cite{Jiang2006b}.Reprinted with permission from App. Phys. Lett. 89, 011908 (2006). Copyright 2006, American Institute of Physics. Bottom, right: regions of single domain order are large enough (tens-of-micrometer-scale) that the Fourier transform (inset) shows 
well-defined Bragg spots.}
\label{fig:collage_non_volatile}
\end{figure}
Unlike the experiments shown in Fig.~\ref{fig:spin coating3} (but like the ones shown in Fig.~\ref{fig:spin coating1}), the first spin coating experiments were carried out in a non-volatile, polymerizable liquid, where the structures could be preserved $via$ ultraviolet curing. In this case, there is thinning of the suspension, but the capillary forces due to drying are absent. 
Jiang and McFarland \cite{Jiang2004} reported 
wafer-size fabrication of colloidal crystals using spin coating of colloidal particles in a non-volatile solvent. 
In their experiment, they spin coated a suspension of silica spherical particles in ethoxylated trimethylolpropane
triacrylate (ETPTA) monomer on a centimeter sized substrate and controlled the thickness of the coating by varying spin speed and spin time. Finally photopolymerization cured the solvent. 
By selectively removing the polymer using an oxygen plasma etcher, they could obtain colloidal crystals with air spaces. By removing silica spheres using hydrofluoric acid aqueous solution, they could get a macroporous polymer. 
Jiang and McFarland produced large area structures (Fig. \ref{fig:collage_non_volatile}, top panel) with six-fold symmetry (left) and demonstrated control of colloidal film thickness (right).
They also found examples with mixed 4- and 6-fold symmetry (Fig. \ref{fig:collage_non_volatile}, bottom left).

\begin{figure}[h]
\centering
\includegraphics[width=80mm]{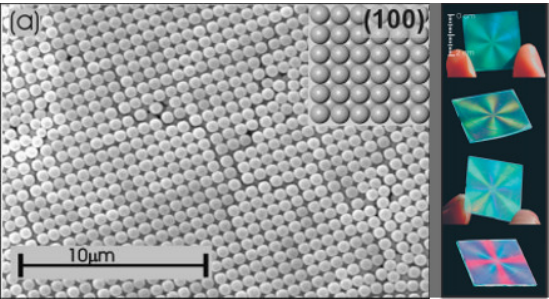} 
\caption{Spin coating in volatile solvents.
Left: Scanning electron micrograph of the top view of a spin coated film where the solvent evaporates completely, which exhibits 4-fold symmetry, in particular the (100) face of an fcc crystal domain. Right: Dried colloidal crystals exhibiting 6-fold and 4-fold symmetry under different spin coating conditions. Adapted from Mihi
{\it et al.}~\cite{Mihi2006}. Reprinted with permission. Copyright 2006, John Wiley and Sons.
}
\label{fig:collage_volatile}
\end{figure}

Mihi {\it et al.} reported the production of colloidal crystals by spin coating using a mixture of volatile solvents that evaporates during spin coating, leaving a colloidal crystal that needs no further processing \cite{Mihi2006}. They used different mixtures of ethanol, distilled water, and ethylene glycol as a solvent for micrometer-range silica spheres. The thickness of the resulting colloidal crystal could be controlled by using different proportions of each solvent in the mixture or by changing the spinning speed.  They also observed that the proportions of the mixture determined the symmetry of the top plane of the colloidal crystal (4-fold in Fig.~\ref{fig:collage_volatile}, left panel), with both 4-fold and 6-fold symmetry being observed (Fig.~\ref{fig:collage_volatile}, right panel).

Two features are common to all the experiments described thus far. First, for both volatile and non-volatile solvents, global (cm-scale) 4- or 6-fold symmetry coincides with local ($\mu$m-scale) symmetries of packing as observed by scanning electron microscopy. Second, the dynamical structures from volatile and non-volatile solvents are very similar. 
Structures produced with a polymeric fluid that does not evaporate (Fig.~\ref{fig:collage_non_volatile}, top left and bottom left) correspond remarkably well to those in Fig.~\ref{fig:spin coating3}(b and c) in the high-speed dynamics experiments. Similarly, the structures observed after drying from the volatile solvent (with either 4- or 6-fold symmetry in Fig.~\ref{fig:collage_volatile}, right) correspond very well to the 4- or 6-fold symmetric dried structures observed at different rotation rates from the dynamical experiments (Fig.~\ref{fig:spin coating3}e,f).
\red{In some other works\cite{Prevo2007}, it is shown that one can control the crystalline structure by controlling the thickness. However for evaporative spincoating such a dependence is not reported. One can control final thickness in both volatile and non-volatile solvents by controlling the spinning speed. Nevertheless, Arcos {\it et al.}\cite{Arcos2008} reported 4-fold symmetry in a wide range of spinning speeds with acetone as solvent and 6-fold symmetry with ethanol as solvent, suggesting some other material parameters, such as the contact angles, are also important.}

\subsection{Nature of the orientational order and the colored patterns}
\begin{figure}[h]
\centering
\includegraphics[width=75mm]{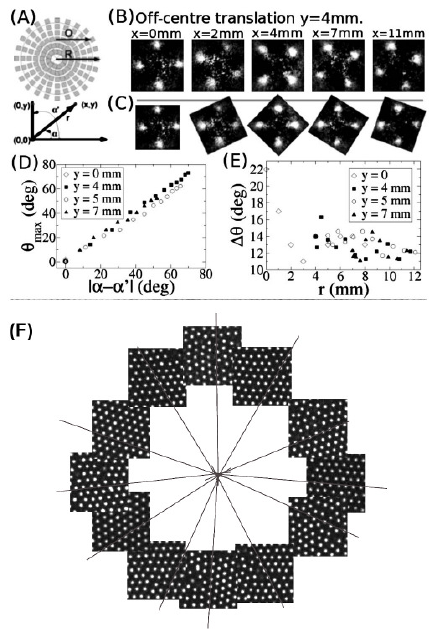} 
\caption{(A) Proposed structure of the spin coated crystal as a polycrystal where different domains 
are orientationally correlated. (B,C) Diffraction patterns obtained at 1 mm positional intervals using a 405 nm laser and (D) the angular correlations of these domains demonstrate the the orientation of the crystalline domains is consistent with the proposed structure. (E) The domain angular dispersion is roughly unchanged so long as one is not too close to the centre of spinning\cite{Arcos2008}. Reprinted with permission from ref. 6. Copyright (2008) by the American Physical Society. http://http://pre.aps.org/abstract/PRE/v77/i5/e050402
(F) Real-space confocal micrographs showing orientationally correlated structure \cite{YuLing2007thesis}. Adapted from Wu \cite{YuLing2007thesis}, with permission from the author.}
\label{fig:diffraction}
\end{figure}
In this section, we address the relationship between local and global symmetries in colloid spin coating. 
A striking feature of the spin coated colloidal films (including the representative examples shown in Figures \ref{fig:collage_non_volatile} and \ref{fig:collage_volatile}) is that there are bright arms with either 4-fold or 6-fold symmetry when the film is viewed under diffuse white light conditions \cite{Arcos2008, Jiang2004}. These arms do not rotate when the sample is rotated; the film has global azimuthal symmetry.
In other words, despite the fact that the local crystalline structure breaks azimuthal symmetry with respect to the center of spinning, the film as a whole does not break azimuthal symmetry.

A picture for how the angular correlations of these local 4- or 6-fold structures resulted in macroscopic azimuthal symmetry was provided by Arcos {\it et al.}\cite{Arcos2008}, and is summarized here in Fig. \ref{fig:diffraction}. By displacing a tightly focused 405 nm laser beam on the colloidal film along an off-center translation $O$, as shown in Fig. \ref{fig:diffraction}(A), the diffraction spots rotate (Fig. \ref{fig:diffraction}(B)). In contrast, a radial translation $R$, (depicted in Fig. \ref{fig:diffraction}(A)) does not rotate the diffraction pattern, as shown in Fig. \ref{fig:diffraction}(C). This observation shows, first, that the crystalline domains are large enough that a mm-diameter laser beam does not see a powder pattern. Second, the rotation of the diffraction pattern demonstrates that the domain orientation undergoes continuous macroscopic rotation on length scales much larger than the lattice spacing. Thus, the colloidal thin film is an orientationally correlated polycrystal (OCP) wherein the crystalline domains are radially arranged with respect to the center of the film \cite{Arcos2008}. Plots in Fig. \ref{fig:diffraction}(D,E) provide more quantitative verification: there is long-ranged orientational correlation between domains over several mm (Fig. \ref{fig:diffraction}(D)) with an angular dispersion of about 12 - 14$^\circ$ (Fig. \ref{fig:diffraction}(E)). Independently, the same picture has been constructed in real space (Fig. \ref{fig:diffraction}(F)) by obtaining confocal micrographs at different orientations, some radial distance from the centre of spinning \cite{YuLing2007thesis}.
There is also direct evidence from cross-sectional images \cite{Mihi2006, Vermolen2008thesis, Arcos2008} as well as scanning electron microscopy, light microscopy, and atomic force microscopy images of the surfaces of spin coated crystals \cite{Vermolen2008thesis, Arcos2008} that the packing of the colloidal spheres can be described locally by an fcc structure.

\begin{figure}[h]
\centering
\includegraphics[scale=0.25]{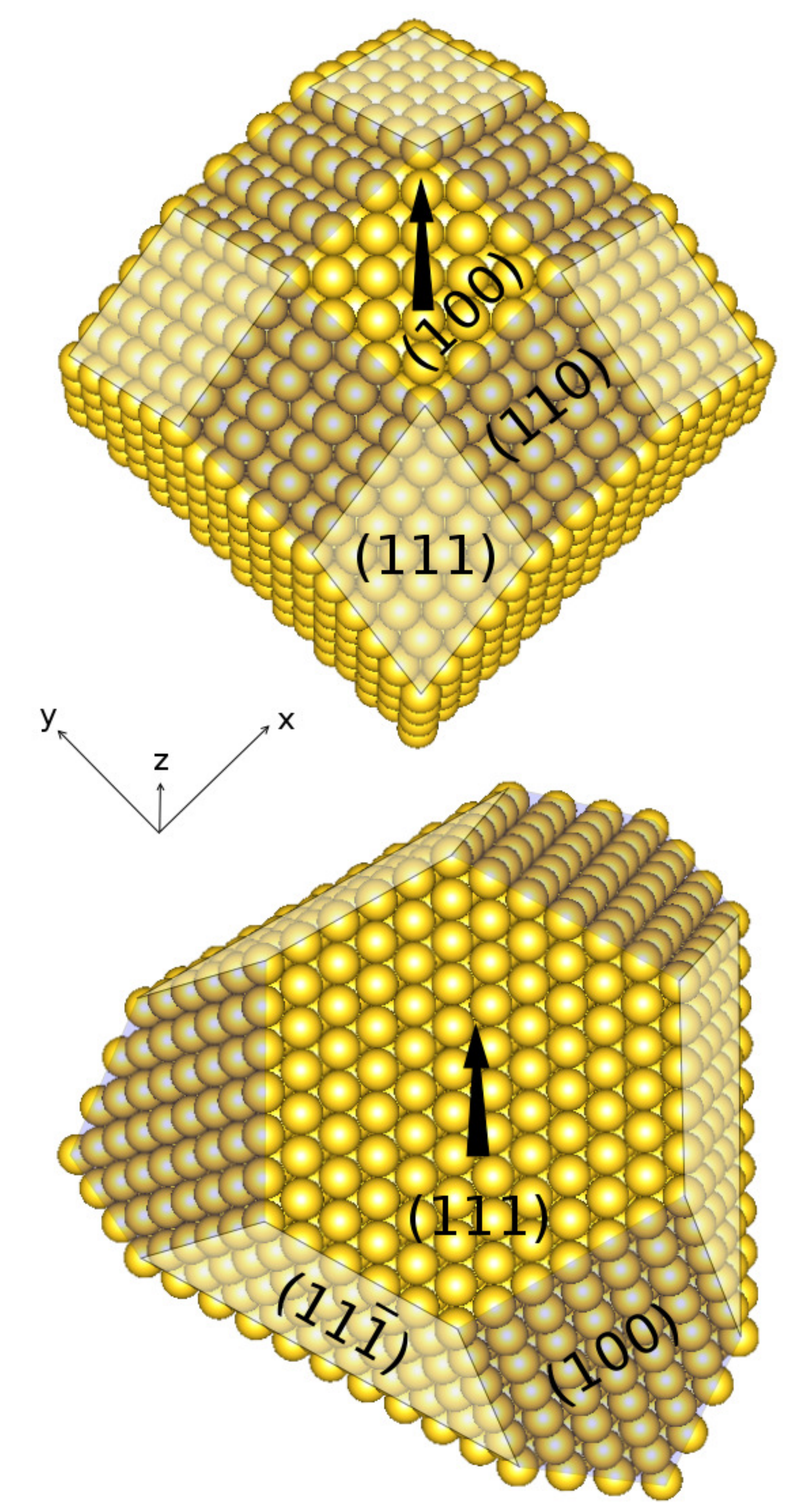} 
\caption{The 4-fold and 6-fold symmetries observed in spin coated colloidal crystals arises from Bragg reflections from different faces of an fcc structure. With the z direction corresponding to the ``top'' view of spin coated structures, fcc structures with (100) and (111) faces parallel to the substrate give rise to 4- and 6-arm patterns
respectively. The global 4 or 6 arm symmetry is believed to arise from Bragg reflections from light incident at oblique angles. In particular, the 4-fold symmetry of the (111) and (110) planes (top) leads to the 4 arm patterns, and the 3-fold symmetry of the (111) and (100) planes (bottom) leads to the 6 arm patterns; see text.}
\label{fig:cuts}
\end{figure}

Understanding the local and orientational arrangement of colloids in the crystal can explain the origin of the symmetric structural colors \cite{Vermolen2008thesis}. The proposed mechanism for the appearance of colored arms on the sample is that the light \kp{diffracts from} Bragg planes (which resembles a specular reflection) \cite{Ruhl2004}.  Figure \ref{fig:cuts} shows an fcc structure with (100) plane on top that is cut by (111) and (110) planes. Notice that it is possible to view each family of planes ((111) and (110)) by rotating this crystal about an axis normal to the (100) plane in 45 degree increments. The idea is that, for four specific orientations of the spin coated crystal, the (111) planes reflect visible light back to the observer,  while (110) planes do the same for the four other positions that lie in between the former four. \red{To be more precise, in the case of samples showing 4-fold symmetry, the plane parallel to the substrate can be either (100) or (200), because 2n+1 (n $\ge$ 1) layers are needed in order to have (100) on the top, which is not necessarily the case.} The net effect is the appearance of two right-angle crosses that are rotated by 45 degrees relative to each other.

A similar argument can be applied to samples with a (111) surface presented, which results in 6-fold symmetric coloured arms. Vermolen has noted \cite{Vermolen2008thesis} that the existence of 6-fold rather than 3-fold symmetry indicates that the stacking of spin coated crystals is twinned. 

The first attempts to understand what experimental parameters dictate the appearance of the 4-fold or 6-fold symmetries were based on fluid dynamical treatments of the solvent and focused on their likelihood to nucleate different faces ((100) for 4-fold and (111) for 6-fold) of a close-packed (fcc) structure \cite{Ishii2007}. However, this still remains an open question.

\section{New Directions}

\subsection{Customized spin coating with external fields}

\subsubsection{Electric fields.~~}
Although spin coating offers reproducibility and robustness for producing polycrystalline colloidal films, it is clear from the discussion in the previous section that this method is incapable of delivering monodomain, defect-free crystals. Axial symmetry from the spinning arranges the microscopic domains of colloidal particles in an orientationally correlated fashion, where microscopic domains have short range positional order and long range orientational order \cite{Arcos2008,Giuliani2010}. Symmetry breaking mechanisms might provide important clues for obtaining crystallites in a privileged direction. Since earlier work has identified key variables that affect colloidal crystallization during spin coating, there are several good starting points for learning how to influence fluid flow and evaporation. 

\begin{figure}[ht]
\centering
\includegraphics[width=4.1cm]{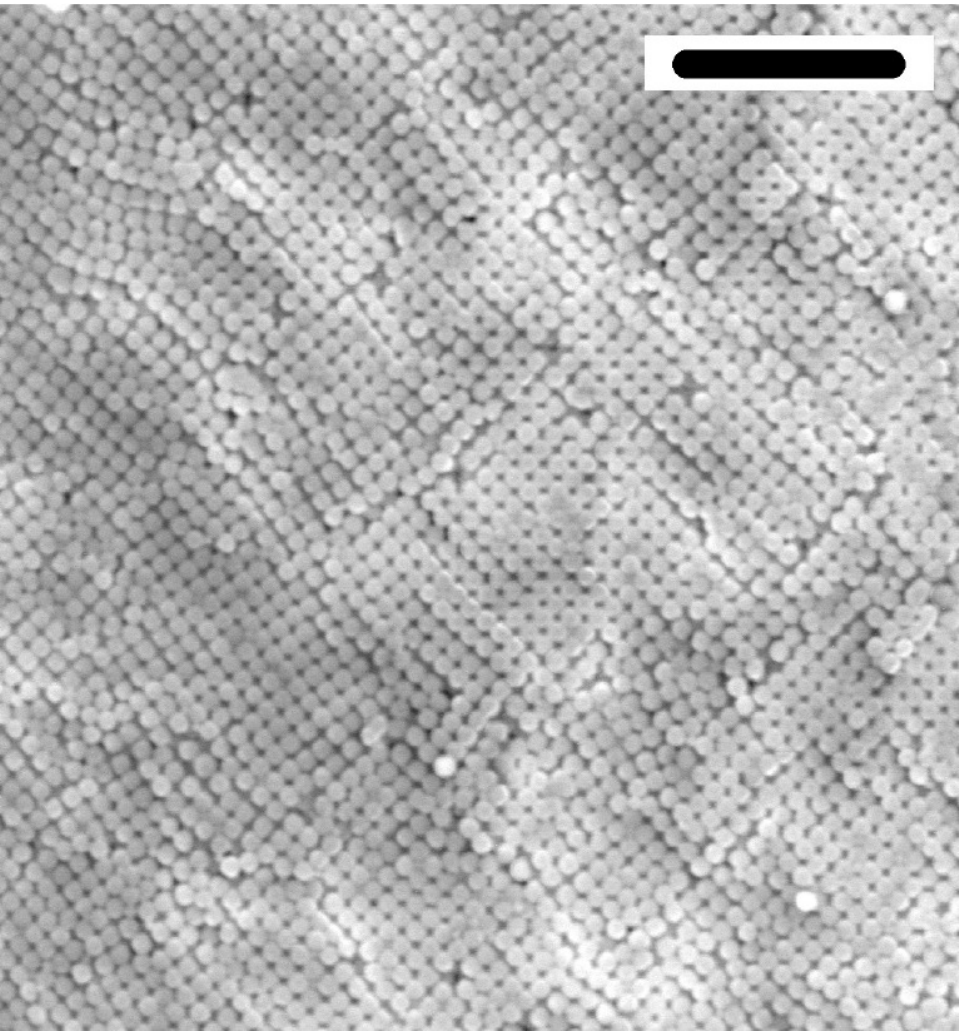}\quad\includegraphics[width=4.1cm]{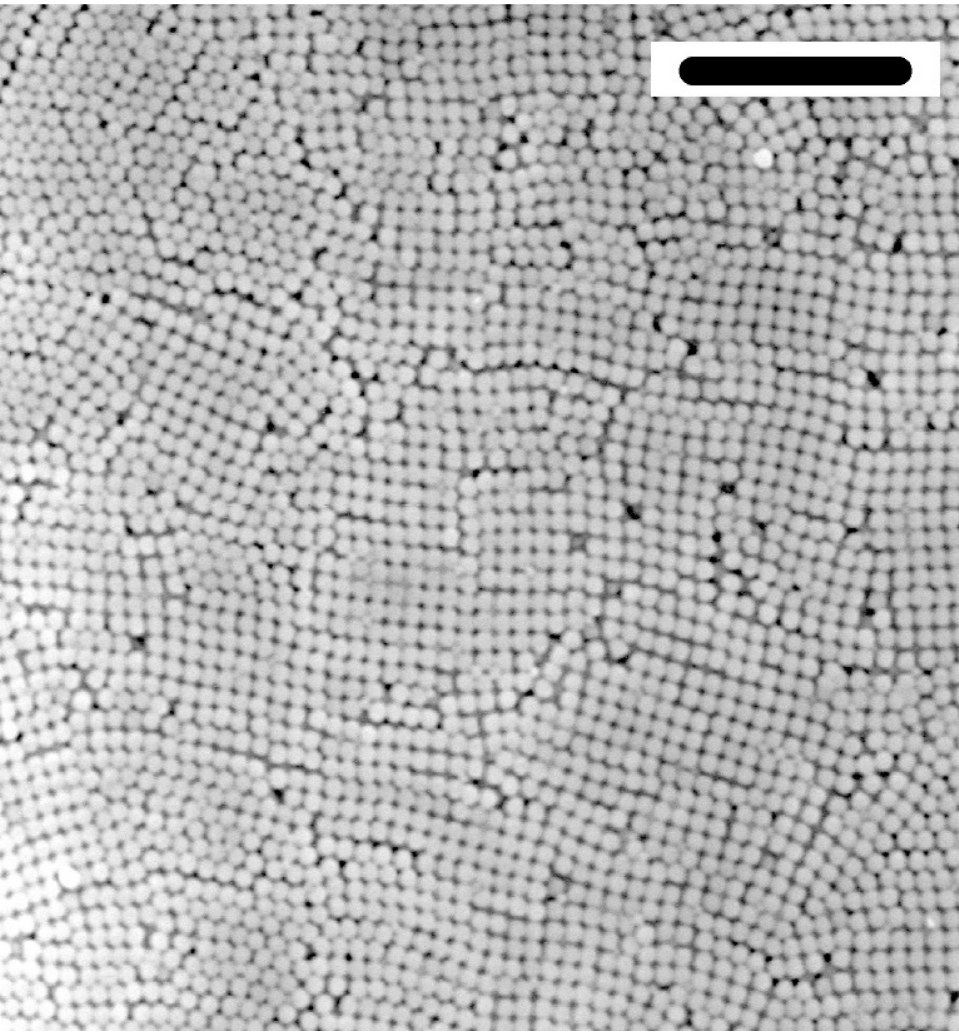}
\caption{Scanning electron micrographs in the absence and in the presence of electric field, left and right respectively. (right)~Field strength 0.95~kV/mm and frequency 3~kHz. 
Left: The dominant domain orientation at zero field (obtained via individual particle tracking methods that are optimized for crowded particle features \cite{mcdonald2012}) is along the radial direction, which is about 45$^\circ$ from the vertical. Right: The dominant domain orientation is along the field direction (vertical) and not in the radial direction (which is 49 $^\circ$ from vertical). The scale bars are 5$\mu$m\cite{bartlett_langmuir12}.(Reprinted from Langmuir 2012, 28, 3067−3070. Copyright 2012 American Chemical Society.) }
  \label{fig:sem}
\end{figure}

Recently, it was demonstrated that the application of a nonuniform electric field while spin coating affects the hydrodynamic flows through dispersion--air dielectric contrast \cite{bartlett_langmuir12}. By arranging the alternating field direction to be fixed in the rotating frame, the axial symmetry from spinning is broken. In the absence of an external field, the colloidal crystals show iridescence with four-fold or six-fold symmetry that is a manifestation of orientationally correlated microscopic domains. The electrode geometry on the substrate assists in symmetry breaking, once the electric field is applied, by directing colloidal deposits along predefined directions. The application of an electric field also changes the net domain orientation, as shown in Fig.~\ref{fig:sem}, because hydrodynamic shear forces and the electrostatic forces compete to orient the domains \cite{bartlett_langmuir12}. These changes in orientation have been assessed with quantitative image processing algorithms \cite{mcdonald2012}.

\subsubsection{Magnetic fields.~~}

The spin coating method has been explored for fabricating data storage devices since the late 1980's \cite{Scriven1988} by utilizing colloidal dispersions of magnetic particles. In this kind of colloidal system, inter-particle magnetic dipole interactions can be affected by applying a magnetic field while spin coating \cite{pichumanimag2011,Pichumani2012,dandapat1999,ilg2005,chen2010mag}.

Experiments that report spin coating in an applied magnetic field are few and recent. \cite{pichumanimag2011, Pichumani2012}. When working with a dilute aqueous colloidal dispersion of superparamagnetic particles, the dominant effect of the applied magnetic field was to change the rheology. There was no evidence of oriented or directed colloid crystallization, but rather particle clusters appeared with sub-monolayer coverages. A model to interpret these results\cite{Pichumani2012} generalized an equation for thickness reported by Cregan {\it et al.} \cite{Cregan2007} by defining a compact equivalent height that accounts for the discrete nature of the particles. Relations between the occupation factor of submonolayer deposits and the dynamic viscosity of the dispersion are consistent with the expected magnetorheological behaviour \cite{Pichumani2012}. Thus, the spin coating method can be used as a rheology probe for rapidly evaporating fluids in open environments, and to measure magnetoviscous effects.

\subsection{Spin coated colloids as templates}

Colloidal crystals based on spherical particles have voids that can be infiltrated with other materials to produce arrays or porous networks. For this reason, colloidal crystals have been widely recognized as 2D and 3D templates, especially by those in the photonics community\cite{Yang2006}. In other applications, patterned magnetic arrays are sought to increase the areal density of hard disks, for magnetic recording read heads, as well as for Magnetoresistive Random Access Memory (MRAM) applications \cite{ross2001, chien2007}.  Although they do not have the perfect crystallinity desired for photonic band gap materials or conventional hard disks, spin coated colloidal crystals can make effective 2D and 3D templates for other related optical and magnetic applications. The impetus for using spin coated templates is in applications wherein the tradeoff between ease of template production outweighs any potential setbacks associated with an imperfect template periodicity. Defects in a colloidal template (vacancies, interstitials, and distortions) can be replicated in an infiltrated material\cite{eagleton2004}. 

True infiltration of templates based on spherical particles leads to interconnected materials. Using electrodeposition, wherein metallic or semiconducting material deposits only on electrically conducting portions of the substrate, the infiltrated material forms a macroporous network \cite{zhukov2006}. Infiltration can also be achieved with liquids\cite{Almanza-Workman2011}. There has also be been a demonstration that one can spin coat twice with different size spheres so that the smaller sphere infiltrate the pores between the larger spheres\cite{xia2008,nandiyanto2011}.  Thinner deposits, less than the height of a single-layer template, can yield hemi-spherical  shapes \cite{Bartlett2003,liu2010}. Bowl-like arrays have also prepared from liquid precursor infiltration \cite{srivastava2006}. As one alternative to infiltration, a single colloid layer can serve as a template by masking to block part of a surface during a vapour phase deposition process, such as sputtering or thermal evaporation. Another avenue is to use the entire sphere as a template to produce a series of porous, interconnected shells\cite{moon2009}.

\subsubsection{Optical applications~~}

The complex shapes that appear in templates prepared by infiltration have been targeted as substrates for surface enhanced Raman spectroscopy (SERS)\cite{liu2010,heo2012}. These patterned, optically active gold substrates have been proposed to be used as biological sensors\cite{Yang2006}.

Spherical colloids have also been used as masks to produce nanohole arrays that can increase the photoluminescence intensity of light-emitting devices such as SiN\cite{oh2010}. Jiang \textit{et al.}\  have used colloidal templates to produce coatings that are both anti-reflective and superhydrophobic\cite{min2008}, as well as half-shell metallic arrays as SERS substrates \cite{liu2010}.

\subsubsection{Magnetic applications~~}

Although most current implementations of magnetic arrays in technological applications favor reliable long range order, it is likely that higher density arrays will require individual device mapping to identify the precise positions of each data storage bit. Thus, it is conceivable that magnetic arrays produced from spin coated templates could be useful in functional devices. To make a useful patterned magnetic array, there are a range of geometric and material parameters that must be balanced against technological demands for higher data densities. The most important consideration is that each element should consist of a single, stable magnetic domain that can be switched between distinct magnetic states. It is desirable for all elements in an array to have a uniform switching field, and for individual elements to be stable with respect to the fields generated by nearby elements.  

Most infiltration studies for magnetic applications have focused on Ni and Co metals and alloys\cite{jiang2006,Arcos2008}, and these metals are typically infiltrated by electrochemical deposition \cite{zhukov2006,napolskii2007,eagleton2004, Bartlett2004}. This includes one
proof-or-principle infiltration of Co into spin coated silica colloidal films \cite{Arcos2008}. In all cases, the templating has a profound effect on the magnetic hysteresis response\cite{Bartlett2003,Liu2006a}. This is expected since magnetic hysteresis effects are not an intrinsic material property and depend entirely on the grain and domain structure of the material. The interconnecting necks in these infiltrated materials have minimum widths that are typically less than a few hundred nanometers, which is below the threshold for single magnetic domains in Ni, Fe, and Co. Because these constrictions qualitatively change the magnetic response of the metal in those ares, the neck regions may contribute significantly to the collective magnetic hysteresis response of the infiltrate material.

Many questions relating the magnetization behaviors of individual magnetic elements prepared $via$ colloidal templates have not yet been adequately addressed using tools such as magnetic force microscopy \cite{zhu2003}. There is also an ongoing need to understand the relationship between collective magnetic properties of arrays of magnetic colloids as a function of template spacing, element shape, and disorder or defect concentration. This has ramifications for the switching field values and its uniformity throughout a patterned magnetic array. In this respect, micromagnetic simulations are playing a critical role to bridge the gap between theory and experiment \cite{kim2005}.

\section{Conclusions}

Spin coated crystals exhibit a variety of structures that are intermediate between perfect order and disorder. 
While fundamental challenges persist in the control of the crystallinity of the resulting films, the primary materials science advantage offered by spin coating is a highly reproducible control of thickness. Spin coated crystals serve as a rich model system for exploring the fundamentals of crystallization in confined environments, and as test cases for assessing polycrystalline domains and defect-rich crystals. The symmetry transitions that occur during the spin coating are not yet understood in detail, and we encourage further study of these phenomena, especially $via$ simulation. 

\section{Acknowledgements}
This work was supported by the National Science and Engineering Research Council of Canada (NSERC) and Spanish MEC (Grant No. FIS2011-24642).  M.P. acknowledges financial support from the “Asociaci{\'o}n de Amigos de la Universidad de Navarra”.




\footnotetext{\textit{$^{a}$~Department of Physics and Applied Mathematics, University of Navarra, Pamplona, Spain. E-mail: wens@unav.es}}
\footnotetext{\textit{$^{b}$~Department of Physics and Physical Oceanography, Memorial University of Newfoundland, St. John's, NL A1B3X7, Canada. E-mail: ayethiraj@mun.ca}}



\providecommand*{\mcitethebibliography}{\thebibliography}
\csname @ifundefined\endcsname{endmcitethebibliography}
{\let\endmcitethebibliography\endthebibliography}{}

\end{document}